\documentclass[10pt,aps,prd,twocolumn,superscriptaddress,nofootinbib,floatfix]{revtex4-2}
\usepackage{graphicx}
\usepackage{amsmath}
\usepackage{amssymb}
\usepackage{bm}
\usepackage[colorlinks=true,citecolor=blue,linkcolor=blue,urlcolor=blue]{hyperref}

\newcommand{\dd}{\mathrm{d}}
\newcommand{\Msun}{M_\odot}

\begin{document}

\title{Gravitational-wave signatures of spinning particles around rotating hairy black holes}

\author{Surojit Dalui}
\email{surojitdalui@shu.edu.cn, surojitdalui003@gmail.com}
\affiliation{Department of Physics, Shanghai University, 99 Shangda Road, Shanghai 200444, China}
\affiliation{Yukawa Institute for Theoretical Physics, Kyoto University, Kyoto 606-8502, Japan}
\author{Xian-Hui Ge}
\email{gexh@shu.edu.cn}
\affiliation{Department of Physics, Shanghai University, 99 Shangda Road, Shanghai 200444, China}

\date{\today}

\begin{abstract}
We study how the spin of a compact secondary and a gravitational-decoupling deformation of the Kerr
geometry are imprinted on gravitational waves from strong-field eccentric motion around a rotating black
hole. The orbit is evolved with the Mathisson--Papapetrou--Dixon equations under the Tulczyjew spin
supplementary condition, and the radiation is computed with the numerical-kludge method. The deformation
describes a black hole only below a critical curve in the plane of the two hair parameters. The spin of
the secondary pulls the pericenter inward and, above a critical value, carries the particle into the black
hole. The hair raises this capture threshold, from $S\simeq0.66$ in Kerr to $0.9$ and above close to the
horizon boundary, because it reduces the effective mass near the black hole. The emitted signal remains a
discrete comb of orbital harmonics: the spin strengthens the pericenter bursts and shifts power toward
higher harmonics, while both the spin and the hair leave their main imprint through an accumulated
dephasing. The imprint of the hair is controlled by the parameter that sets the radial reach of the
deformation, disappears in the Kerr-like limit of the family, and is amplified by the spin of the secondary.
For a source at $2$~Gpc with $M=10^6M_\odot$ and $\mu=10M_\odot$, a $6.8$~h segment of the signal carries a
signal-to-noise ratio of about $1.4$ in LISA, $2.7$ in Taiji and $0.9$ in TianQin. Within the same segment the more
extended deformations become distinguishable from Kerr, first in Taiji; since this holds for a nonspinning
secondary, it applies directly to astrophysical EMRIs. The spin-induced frequency shifts are linear in the
spin, so the large spins used here are a magnified illustration of a dephasing that, for realistic
secondaries, accumulates to a phase of order a radian over an inspiral.
\end{abstract}

\maketitle

\section{Introduction}
\label{sec:intro}

The observation of gravitational waves (GWs) has transformed compact-object dynamics from a largely
theoretical subject into an observational probe of the strong-field regime of
gravity~\cite{Abbott2016,Abbott2017,Abbott2019GWTC1,Abbott2021GWTC2,Abbott2023GWTC3,Abbott2024GWTC21}.
Horizon-scale imaging by the Event Horizon Telescope has opened a complementary electromagnetic window
on the immediate environment of supermassive compact objects~\cite{EHT2019a,EHT2019f,EHT2019d,EHT2022}.
Together, these developments make it increasingly meaningful to ask not only whether a compact object is
compatible with a Kerr black hole, but also how small departures from Kerr could appear in observables.

Extreme-mass-ratio inspirals (EMRIs) are especially attractive in this respect. A stellar-mass compact
object can spend a very large number of orbital cycles in the strong gravitational field of a massive
black hole, encoding detailed information about the background spacetime in the phase and harmonic
structure of the emitted radiation~\cite{AmaroSeoane2017,Babak2017,Barack2019,Hughes2000,Hughes2005,DrascoHughes2004,DrascoHughes2006}.
This is one of the principal motivations for space-based detectors such as
LISA~\cite{AmaroSeoane2017,RCL2019}, TianQin~\cite{Luo2016,Mei2021,Li2025} and
Taiji~\cite{HuWu2017,Liu2023}. The long duration of an EMRI is an advantage, but it also makes waveform
modelling unusually demanding: effects that are perturbatively small over one orbit can accumulate into
measurable phase differences over many cycles~\cite{BarackCutler2004,BarackCutler2007}.

Most EMRI studies begin with a Kerr background, as expected from the vacuum no-hair paradigm of general
relativity. Nevertheless, there are several reasons to study controlled deformations of Kerr. Compact
objects with scalar, vector or Proca hair, scalarized solutions, regular black holes, effective matter
distributions and horizonless mimickers can all produce strong-field geometries that differ from the
Kerr solution while remaining sufficiently compact to imitate some of its observational
properties~\cite{Liebling2023,Mazur2023,CardosoPani2019,Silva2018,Doneva2018,Herdeiro2014,Fan2016,Herdeiro2016,Babichev2017,AyonBeato1998}.
The question relevant for GW astronomy is then not simply whether such alternatives exist, but which
pieces of their geometry can survive into a measurable orbital or waveform signature.

In this work we use a rotating hairy black hole generated through gravitational decoupling (GD). The GD
approach provides a convenient way to add an independently conserved source to a known seed geometry
and to keep track of the resulting deformation~\cite{Ovalle2017,Ovalle2019,Ovalle2018MGD,OvalleCasadio2021}.
Its extension to rotating systems yields a Kerr-like metric with a radially varying effective mass
function, without requiring the Newman--Janis algorithm~\cite{Contreras2021}. Thermodynamic, horizon and
phenomenological properties of related hairy geometries have been explored in a number of
settings~\cite{Mahapatra2023,Ovalle2021KdS,Ovalle2022,Dasgupta2025}. A useful feature of the present
geometry is that the deformation is localized in radius. Consequently, the ``amount of hair'' sampled by
an orbit depends on where that orbit actually travels; it cannot in general be summarized by a single
monotonic strength parameter.

There is a second ingredient that becomes important for compact secondaries: intrinsic spin. The motion
of a spinning body is not geodesic even when its mass is small enough to neglect its backreaction on the
background. At pole--dipole order it is described by the Mathisson--Papapetrou--Dixon (MPD)
equations~\cite{Mathisson1937,Mathisson2010,Papapetrou1951,Dixon1970}. The spin couples directly to
curvature, producing a force that changes the orbital frequencies, turning points and precession.
Spinning-particle motion has been studied extensively in Schwarzschild and Kerr spacetimes, including
its gravitational radiation, its integrability properties, resonances and chaotic
regimes~\cite{Tanaka1996,Saijo1998,SuzukiMaeda1997,SuzukiMaeda1998,Hartl2003a,Hartl2003b,Han2008,Han2010,LukesSeyrich2016,Zelenka2020,Tod1976,Semerak1999}.
Hamiltonian formulations and connections with effective-one-body dynamics provide complementary
descriptions~\cite{Barausse2009,Witzany2019a,Witzany2019b,Gonzo2024,Piovano2025,Albertini2025}.

The combination of a non-Kerr background and a spinning secondary is physically interesting because the
two effects do not simply add. The MPD force contains the background curvature explicitly, so changing
the geometry changes the force experienced by the spin; at the same time, changing the spin moves the
orbit into a different region of that geometry. Studies in regular, quantum-corrected,
dark-matter-dressed and other deformed black-hole backgrounds have already shown that spinning-particle
dynamics can be highly sensitive to the detailed strong-field
structure~\cite{Zhang2020,Ladino2022,Tan2025,Mannobova2025,Umarov2025,Jumaniyozov2025a}.

This sensitivity to the near-horizon geometry is not specific to spinning bodies. Particle motion close
to a horizon generically becomes chaotic~\cite{Hashimoto2017,Dalui2019}; even a Killing horizon in flat
spacetime, as seen by a uniformly accelerated observer, is enough to induce chaos in an otherwise integrable
system~\cite{Dalui2020IJMPA}. The effect persists when quantum corrections to the geometry are included,
which make chaos set in at lower energies~\cite{Bera2022}, in $f(R)$ theories of gravity~\cite{Das2024}, and
in Schwarzschild--de~Sitter spacetime, where it emerges as the black-hole and cosmological horizons approach
each other~\cite{Dalui2026SdS}. The local instability of radial motion near the horizon, present in both
static and Kerr black holes, has also been identified as the origin of the thermal character of the
horizon~\cite{Dalui2020PRD,Dalui2022Kerr}. The rate of this instability, the Lyapunov exponent, is governed
by the surface gravity of the black hole, the quantity that also sets the universal bound on
chaos~\cite{Maldacena2016}. This bound can be saturated or even violated when the particle carries charge
and angular momentum~\cite{Lei2022}, in the presence of quasi-topological electromagnetism~\cite{Lei2021},
or near charged black branes with hyperscaling violation~\cite{Lei2023}; its violation has been linked to
the thermodynamic stability of the black hole~\cite{Lei2024}, and extremal black holes, which have zero
temperature, still show chaotic motion with a positive Lyapunov exponent~\cite{Dalui2026extremal}. Our companion
analysis of the same rotating hairy geometry found that the finite-time instability of MPD trajectories
is organized in localized regions of spin--hair parameter space rather than increasing monotonically
with either parameter alone~\cite{DaluiGe2026}. This naturally raises the observable question addressed
here: what does the dynamics of a spinning secondary in this geometry do to the gravitational waveform?

A fully self-consistent answer would require solving the perturbation equations of the rotating hairy
spacetime with a spinning source and including radiation reaction. That is beyond the scope of the
present work. Instead we use the numerical-kludge (NK) strategy, which has a long history in EMRI
waveform studies~\cite{RuffiniSasaki1981,Sasaki1983,Babak2007,Press1977,Thorne1980,PoissonWill2014} and
which we have recently applied to EMRIs around black holes embedded in dark-matter
halos~\cite{Das2025a,Das2025b}. The philosophy is simple: the trajectory is computed relativistically in
the curved spacetime, while the emitted radiation is estimated by mapping that trajectory to an auxiliary
flat-space source and applying a multipolar radiation formula. The NK method does not provide precision
waveforms for parameter estimation, but it is well suited to identifying robust changes in phasing and
harmonic content across a family of strong-field orbits. Similar strategies have been used to connect
complicated or chaotic orbital dynamics with GW signatures in other
systems~\cite{SuzukiMaeda2000,Kiuchi2004,Kokubun1998,Kiuchi2007,Yang2025,Yuan2026}.

Our goal is therefore deliberately narrower than precision EMRI modelling. We ask which features of the
waveform can be traced to the secondary spin and to the radial structure of the hair, and whether they
are large enough to be seen by a space-based detector. We keep the requested orbital geometry fixed and
vary the particle spin and the two deformation parameters. Once the MPD constraints and Killing
integrals are imposed, changing the spin does not merely multiply the same force along the same orbit:
it changes the self-consistent initial momentum, the spin tensor and the values of the conserved energy
and angular momentum. The resulting trajectories therefore have different empirical turning points even
though they originate from the same requested $(r_p,e,\iota)$ labels.

Our main findings are as follows. The deformation admits a horizon only below a critical curve in the
plane of the hair parameters. The spin of the secondary pulls the pericenter inward and, above a critical
value, carries the particle into the black hole; the hair raises this capture threshold, most strongly just
above the boundary where the horizon disappears. The waves stay a discrete superposition of orbital
harmonics throughout, so the observable effect of the spin and the hair is an accumulated dephasing
together with a redistribution of power among the harmonics. The hair acts almost entirely through the
parameter that controls the radial reach of the deformation, its Kerr-like limit reproduces the Kerr
waveforms exactly, and its imprint grows with the spin of the secondary. We quantify detectability through
the signal-to-noise ratio in LISA, TianQin and Taiji, and distinguishability through the noise-weighted
overlap between waveforms.

The paper is organized as follows. Section~\ref{sec:metric} introduces the rotating hairy background and
its horizon domain. Section~\ref{sec:mpd} summarizes the MPD dynamics and the meaning of the spin
parameter. Section~\ref{sec:nk} describes the orbital setup and the NK waveform construction.
Section~\ref{sec:results} presents the orbital dynamics, the waveforms in the time and frequency domains,
the imprint of the hair and the detectability of the signals. Section~\ref{sec:disc} discusses the
physical interpretation and the limitations of the analysis, and Sec.~\ref{sec:concl} concludes. The
numerical evolution and its validation are described in Appendix~\ref{app:num}. We use geometrized units
$G=c=1$ and the metric signature $(-,+,+,+)$.

\section{Rotating hairy black-hole background}
\label{sec:metric}

We consider the axisymmetric GD geometry of Ref.~\cite{Contreras2021},
\begin{align}
\dd s^2 =& -\left(1-\frac{2r\tilde m(r)}{\Sigma}\right)\dd t^2
          -\frac{4ar\tilde m(r)\sin^2\theta}{\Sigma}\,\dd t\,\dd\phi \nonumber\\
         &+\frac{\Sigma}{\Delta}\,\dd r^2+\Sigma\,\dd\theta^2
          +\frac{\mathcal{A}\sin^2\theta}{\Sigma}\,\dd\phi^2 ,
\label{eq:metric}
\end{align}
with
\begin{align}
\Sigma &= r^2+a^2\cos^2\theta, \\
\Delta &= r^2+a^2-2r\tilde m(r), \\
\mathcal{A} &= (r^2+a^2)^2-a^2\Delta\sin^2\theta ,
\end{align}
and
\begin{equation}
\tilde m(r) = M-\frac{\alpha r}{2}\exp\!\left(-\frac{r}{M-\beta/2}\right).
\label{eq:mtilde}
\end{equation}
The Kerr limit is recovered for $\alpha=0$. The parameter $\alpha$ sets the amplitude of the
deformation, whereas $\beta$ changes its radial localization through $L_\beta=M-\beta/2$; the energy
conditions of the decoupling construction require $\beta<2M$, that is $L_\beta>0$~\cite{Contreras2021}.
This distinction is central to the waveform results below. The correction to the mass function decays as
$e^{-r/L_\beta}$, so for an orbit sampling $r\sim4$--$12M$ changing $L_\beta$ can matter far more than
changing the overall prefactor $\alpha$. At $r=4M$ the exponential factor is about $10^{-3}$ for
$\beta=0.8$ but only about $10^{-9}$ for $\beta=1.6$. Large values of $\beta$ therefore describe a black
hole that is practically Kerr outside the horizon for any $\alpha$, and we use $\beta=1.6$ below as a
Kerr-like reference inside the same family.

\begin{figure*}[t]
\includegraphics[width=\textwidth]{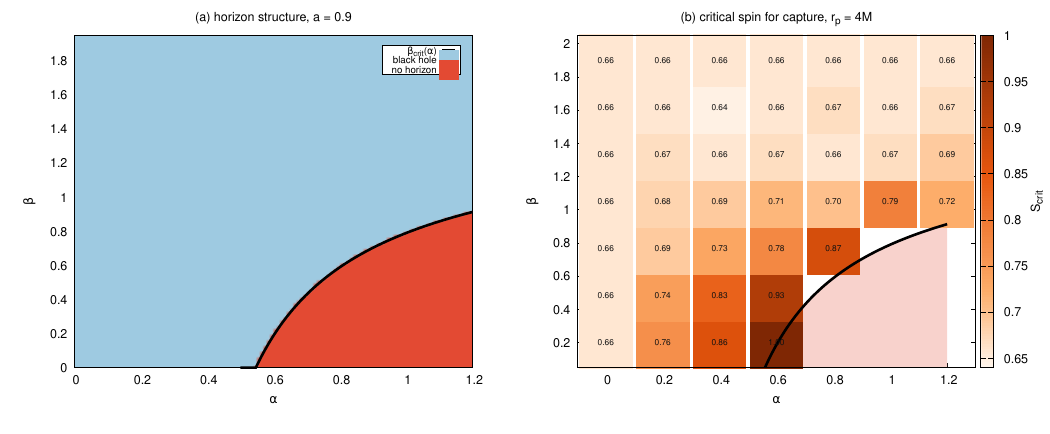}
\caption{(a) Horizon structure of the rotating hairy black hole in the $(\alpha,\beta)$ plane for $a=0.9$.
Blue points describe black holes, red points have no horizon, and the black curve is the critical locus
$\beta_{\rm crit}(\alpha)$ on which the two horizon roots merge. For $\alpha\lesssim0.55$ every $\beta<2M$ gives a
black hole; above this value a minimum $\beta$ is needed. (b) Critical spin $S_{\rm crit}$ above which a
spinning particle on the orbit $r_p=4M$, $e=0.5$, $\iota_{\rm orb}=15^\circ$ is captured within $5000M$, shown
only where a horizon exists (the shaded region has none). In Kerr ($\alpha=0$) and for $\beta\gtrsim1.6$ one finds
$S_{\rm crit}\simeq0.66$. The threshold rises toward small $\beta$ and larger $\alpha$ and is largest next to the
horizon boundary; at $(\alpha,\beta)=(0.6,0.2)$ the orbit is not captured even for $S=1$.}
\label{fig:param}
\end{figure*}

The event horizon is determined by the largest positive root of $\Delta(r)=0$. Not every pair
$(\alpha,\beta)$ produces a black hole, and Fig.~\ref{fig:param} makes this restriction explicit before
any orbital or waveform result is discussed. In Fig.~\ref{fig:param}(a), the blue region contains
horizon roots and therefore represents the black-hole sector, whereas the red region has no horizon. The
black curve, $\beta_{\rm crit}(\alpha)$, is the limiting locus at which the two horizon roots merge. For
$\alpha\lesssim0.55$ the entire displayed range of $\beta$ remains horizon admitting. Once $\alpha$ is
increased beyond this value, a minimum value of $\beta$ is required to retain a horizon: the no-horizon
region first opens near $\beta\simeq0$ and grows steadily upward as $\alpha$ increases, with
$\beta_{\rm crit}\simeq0.19$, $0.43$, $0.59$, $0.70$ and $0.79$ at $\alpha=0.6$, $0.7$, $0.8$, $0.9$ and
$1.0$. Increasing the amplitude parameter $\alpha$ at small $\beta$ therefore does not indefinitely
produce a more strongly deformed black hole; beyond the critical curve it removes the horizon altogether.

The right panel of Fig.~\ref{fig:param} shows a dynamical property of the same black holes, the smallest
spin for which a spinning particle on our reference orbit is captured. We discuss it in
Sec.~\ref{sec:fate}; here we only note that it changes most strongly in the same corner of the plane where
the horizon is about to disappear.

\section{Spinning-particle dynamics}
\label{sec:mpd}

At pole--dipole order the secondary obeys~\cite{Mathisson1937,Papapetrou1951,Dixon1970}
\begin{align}
\frac{Dp^\mu}{\dd\tau} &= -\frac12 R^{\mu}{}_{\nu\alpha\beta}v^\nu S^{\alpha\beta}, \label{eq:mpd1}\\
\frac{DS^{\mu\nu}}{\dd\tau} &= p^\mu v^\nu-p^\nu v^\mu , \label{eq:mpd2}
\end{align}
with the Tulczyjew condition
\begin{equation}
p_\mu S^{\mu\nu}=0 .
\label{eq:ssc}
\end{equation}
Here $v^\mu=\dd x^\mu/\dd\tau$ is the tangent to the worldline. We normalize $p^\mu p_\mu=-\mu^2$ and
describe the spin by the vector $S_\mu=-\frac{1}{2\mu}\epsilon_{\mu\nu\alpha\beta}\,p^\nu S^{\alpha\beta}$,
whose magnitude is conserved, and by the dimensionless spin magnitude
\begin{equation}
S=\frac{S_{\rm phys}}{\mu M}, \qquad S^\mu S_\mu = S_{\rm phys}^2 .
\label{eq:Sdef}
\end{equation}
With the Tulczyjew condition the four-velocity is fixed by the momentum and spin through the exact
relation~\cite{Tod1976,EhlersRudolph1977,Semerak1999}
\begin{equation}
v^\mu=\mathcal N\left[p^\mu+\frac{2S^{\mu\nu}R_{\nu\lambda\kappa\rho}\,p^\lambda S^{\kappa\rho}}
{4\mu^2+R_{\alpha\beta\gamma\delta}S^{\alpha\beta}S^{\gamma\delta}}\right],
\label{eq:vp}
\end{equation}
where $\mathcal N$ is fixed by $v^\mu v_\mu=-1$. We integrate Eqs.~\eqref{eq:mpd1}--\eqref{eq:vp} directly
in terms of the spin tensor $S^{\mu\nu}$, without expanding the velocity relation in the spin and without
passing through dual curvature tensors. The spin vector is used only to specify the initial orientation.
The reasons for this choice, and the checks of the conservation laws, are given in Appendix~\ref{app:num}.
The denominator in Eq.~\eqref{eq:vp} also carries physical information: when $R\,S^2$ becomes comparable to
$\mu^2$ the relation between velocity and momentum degenerates and the pole--dipole description ceases
to be meaningful. For the orbits reported below the denominator never falls below $3.8\mu^2$.

Stationarity and axisymmetry provide the conserved quantities
\begin{align}
E &= -p_t+\frac12 g_{t\mu,\nu}S^{\mu\nu}, \label{eq:E}\\
J_z &= p_\phi-\frac12 g_{\phi\mu,\nu}S^{\mu\nu}. \label{eq:J}
\end{align}
In contrast with Kerr geodesics there is in general no Carter-like third integral, so the orbit is not
reducible to quadratures. When the requested orbital geometry is held fixed and $S$ is changed, the spin
contribution to $E$ and $J_z$ changes as well, so runs with different $S$ are different nonlinear MPD
trajectories rather than one orbit subjected to a progressively stronger copy of the same force. In our
integrations $E$ and $J_z$ are conserved to better than $2\times10^{-9}$ for every bound orbit.

A remark on the size of the spin is in order. For a compact secondary with dimensionless Kerr parameter
$\chi\le1$ one has $S_{\rm phys}\sim\chi\mu^2$ and hence $S\sim\chi\,\mu/M$. For the mass ratio
$\mu/M=10^{-5}$ adopted below this gives $S\lesssim10^{-5}$, and the pole--dipole approximation itself
requires $S\ll1$. None of the values $S=0.05$--$0.8$ used in this paper is therefore an astrophysical EMRI spin; they
are a controlled way of magnifying the spin--curvature coupling, in the same spirit as
Refs.~\cite{SuzukiMaeda1997,Hartl2003a,Zelenka2020,DaluiGe2026}. We verify explicitly in Sec.~\ref{sec:linear} that the spin-induced frequency and phase shifts are
linear in $S$ in the small-spin regime, so that the large-$S$ results reported below can be read as a
magnified illustration of the physical mechanism rather than as a direct astrophysical prediction.

\section{Orbital setup and numerical-kludge waveform}
\label{sec:nk}

We use a common requested orbit with
\begin{equation}
a=0.9M,\quad r_p=4M,\quad e=0.5,\quad \iota_{\rm orb}=15^\circ,
\end{equation}
so that the requested apocenter is $r_a=12M$ and the initial radius is $r_0=(r_p+r_a)/2=8M$. The
initial data are built from the geodesic of the hairy background with these turning points. The
particle starts on the equatorial plane at $r_0$, with the canonical momenta $p_t$ and $p_\phi$ of that
seed geodesic, the radial momentum of the seed at $r_0$, and $p_\theta$ fixed by the mass shell. The
spin vector points along the polar direction of the local orthonormal frame, with its time component
fixed by the Tulczyjew condition, and the spin tensor follows from it. The conserved quantities
$E$ and $J_z$ of Eqs.~\eqref{eq:E} and~\eqref{eq:J} are then determined by this state and differ from the
seed values by the spin terms. The requested parameters therefore label the construction of the initial
data; they are not invariants of the spinning orbit, and the empirical turning points are measured from
the evolved trajectory.

For the radiation we use the NK prescription of Babak \emph{et al.}~\cite{Babak2007}, in the form we
employed in Ref.~\cite{Das2025b}. Over a few orbital periods the energy radiated by the secondary is tiny
compared with its orbital energy, which justifies evolving a conservative orbit and computing the
radiation on top of it~\cite{Hughes2005,Zi2024}. For our parameters the radiation-reaction time estimated
from the Peters formula~\cite{Peters1964}, with semi-major axis $8M$ and $e=0.5$, is
$T_{\rm rr}\sim10^6M$, roughly three months for $M=10^6\Msun$ and $\mu=10\Msun$. The observation window
$T=5000M\simeq6.8$~h used below is shorter by more than two orders of magnitude, and we neglect
radiation reaction throughout.

The Boyer--Lindquist coordinates are mapped to auxiliary Cartesian coordinates,
\begin{equation}
x=r\sin\theta\cos\phi,\quad y=r\sin\theta\sin\phi,\quad z=r\cos\theta ,
\end{equation}
and the Boyer--Lindquist time $t$ is used as the time of the flat space. This is the time measured by a
distant observer, and the orbital frequencies that appear in the signal are frequencies in $t$. It
differs appreciably from the proper time of the particle: along our orbits $\dd t/\dd\tau$ averages to
$1.26$ and approaches $1.7$ at pericenter. In the slow-motion limit the metric perturbation follows from
the quadrupole formula~\cite{Press1977,Thorne1980,PoissonWill2014},
\begin{equation}
h^{ij}=\frac{2\mu}{D_L}\left(a^ix^j+x^ia^j+2v^iv^j\right),
\label{eq:quad}
\end{equation}
where $v^i=\dd x^i/\dd t$ and $a^i=\dd^2x^i/\dd t^2$. We project it onto the detector frame of
Refs.~\cite{Babak2007,Das2025b},
\begin{align}
\mathbf e_X &= (\cos\zeta,\,-\sin\zeta,\,0),\nonumber\\
\mathbf e_Y &= (\cos\iota\sin\zeta,\,\cos\iota\cos\zeta,\,-\sin\iota),
\end{align}
where $\iota$ is the inclination of the line of sight and $\zeta$ the longitude of pericenter, and obtain
the two polarizations
\begin{equation}
h_+=\tfrac12\left(e_X^ie_X^j-e_Y^ie_Y^j\right)h_{ij},\qquad
h_\times=e_X^ie_Y^jh_{ij}.
\label{eq:pol}
\end{equation}
Because both basis vectors are transverse to the line of sight, these combinations pick out the
transverse--traceless part of $h_{ij}$ automatically. We take $\iota=\zeta=\pi/4$, as in
Ref.~\cite{Das2025b}, and adopt $M=10^6\Msun$, $\mu=10\Msun$ and a luminosity distance $D_L=2$~Gpc. The
spin of the secondary enters the waveform only through the trajectory; its own current multipole is of
higher order in $\mu/M$ and is neglected, as is usual in kludge models.

In the frequency domain we use the characteristic strain of each polarization and its combination,
\begin{align}
h_c^{+,\times}(f)&=2f\,|\tilde h_{+,\times}(f)|, \nonumber\\
h_c(f)&=2f\sqrt{|\tilde h_+(f)|^2+|\tilde h_\times(f)|^2},
\label{eq:hc}
\end{align}
computed from the $T=5000M$ record with a Tukey window (taper fraction $0.1$), giving a frequency
resolution $\Delta f=1/T\simeq4\times10^{-5}$~Hz. The NK waveform is not a precision solution of the
perturbation equations of the hairy spacetime. The orbital frequencies, and hence the phasing and
harmonic content of the signal, are inherited exactly from the relativistic trajectory, whereas the
absolute amplitude depends on the flat-space emission formula; in Kerr the NK waveforms overlap with
Teukolsky-based ones at the level of about $95\%$~\cite{Babak2007}. We therefore place most weight on
phasing, harmonic structure and comparisons between configurations treated with the same prescription.

\section{Results}
\label{sec:results}

\subsection{Spin, hair and the fate of the orbit}
\label{sec:fate}

\begin{figure*}[t]
\includegraphics[width=\textwidth]{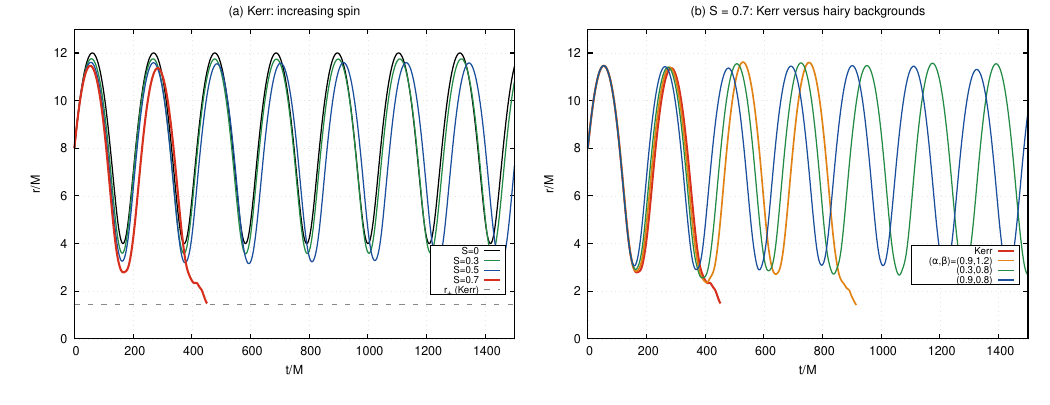}
\caption{Radial motion of the spinning particle for the requested orbit $r_p=4M$, $e=0.5$. (a) Kerr, with
increasing spin. The inner turning point moves inward as $S$ grows, while the outer one stays close to $12M$;
for $S=0.7$ the particle is captured during its second pericenter passage. The dashed line marks the Kerr
horizon. (b) The same spin, $S=0.7$, in different backgrounds. The Kerr orbit is captured at $t\simeq450M$,
the orbit in the $(0.9,1.2)$ background survives about twice as long, and in the two backgrounds with
$\beta=0.8$ the orbit remains bound.}
\label{fig:orbit}
\end{figure*}

The first result is orbital rather than radiative. The spin--curvature force does not simply
nudge the orbit; it moves the pericenter inward, and if the spin is large enough the particle is
carried across the separatrix and falls into the black hole. Figure~\ref{fig:orbit}(a) shows this for
Kerr. All four orbits start from the same point, $r_0=8M$, and follow almost the same path during the
first radial cycle. For $S=0$ the particle oscillates between exactly $4M$ and $12M$, as requested. As
the spin is increased the inner turning point moves inward: it is at $3.57M$ for $S=0.3$ and at
$3.16M$ for $S=0.5$, while the outer turning point moves only slightly, from $12M$ to about $11.6M$.
For $S=0.7$ the second pericenter passage already reaches the plunge region, and the particle falls
into the black hole at $t\simeq450M$.

Figure~\ref{fig:orbit}(b) repeats the experiment at the same spin, $S=0.7$, in different backgrounds,
and shows that the hair changes the outcome. The Kerr orbit plunges, as in panel (a). With
$(\alpha,\beta)=(0.9,1.2)$ the orbit survives about twice as long and plunges at $t\simeq900M$. For the
two backgrounds with $\beta=0.8$ the orbit does not plunge at all and remains bound for the whole
$5000M$ evolution. The physical reason is visible directly in Eq.~\eqref{eq:mtilde}. Close to the black
hole the effective mass $\tilde m(r)$ is smaller than $M$, so the attraction that pulls a spinning
particle over the separatrix is weaker. The smaller $\beta$, the further out this reduction extends,
and the larger the spin needed to capture the particle.

Table~\ref{tab:capture} makes this quantitative by repeating the evolution at four values of the
requested pericenter. In Kerr a spin of $0.3$ is already enough for capture at $r_p=2.5M$ and
$3M$, whereas in the hairy background with $(\alpha,\beta)=(0.9,0.8)$ the same orbits stay bound up to
$S=0.5$ at $r_p=3M$ and up to at least $S=0.7$ at $r_p\ge3.5M$. In all cases the conserved energy is
below unity, so none of these orbits can escape; the only possible change of fate is capture.

The right panel of Fig.~\ref{fig:param} turns this into a map. For every horizon-admitting point
of the $(\alpha,\beta)$ plane we have determined the smallest spin $S_{\rm crit}$ for which an orbit
with $r_p=4M$ is captured within $5000M$. In Kerr, and along the whole column $\beta\gtrsim1.6$ where the
deformation is switched off, $S_{\rm crit}\simeq0.66$. Moving toward small $\beta$ and larger $\alpha$,
the threshold rises steadily, and the largest values are found right next to the critical curve on which
the horizon disappears. There $S_{\rm crit}$ reaches $0.87$--$0.93$, and at $(\alpha,\beta)=(0.6,0.2)$, the point of
our grid closest to the place where the boundary opens, the orbit is not captured even for $S=1$. The capture threshold is therefore
an observable property of the black hole that responds to the hair in the same place where the
geometry is most strongly deformed: just on the black-hole side of the horizon boundary.

\begin{table}[t]
\caption{Smallest spin, among $S=0,0.1,0.3,0.5,0.7$, for which an orbit with requested pericenter
$r_p$ (and $e=0.5$, $\iota_{\rm orb}=15^\circ$) is captured within $5000M$. A dash means that all
these spins remain bound.}
\label{tab:capture}
\begin{ruledtabular}
\begin{tabular}{ccc}
$r_p/M$ & Kerr & $(\alpha,\beta)=(0.9,0.8)$ \\
\colrule
2.5 & 0.3 & 0.5 \\
3.0 & 0.3 & 0.7 \\
3.5 & 0.5 & --- \\
4.0 & 0.7 & --- \\
\end{tabular}
\end{ruledtabular}
\end{table}

For the waveform study we keep $r_p=4M$. There, all spins up to $S=0.5$ are bound in every background
we consider, and in the $(0.9,0.8)$ background the orbit stays bound up to $S=0.8$. This gives a common
set of spins for comparing backgrounds and a wider range of spins for the strongly deformed case.

\subsection{Time-domain waveforms}
\label{sec:time}

\begin{figure*}[t]
\includegraphics[width=\textwidth]{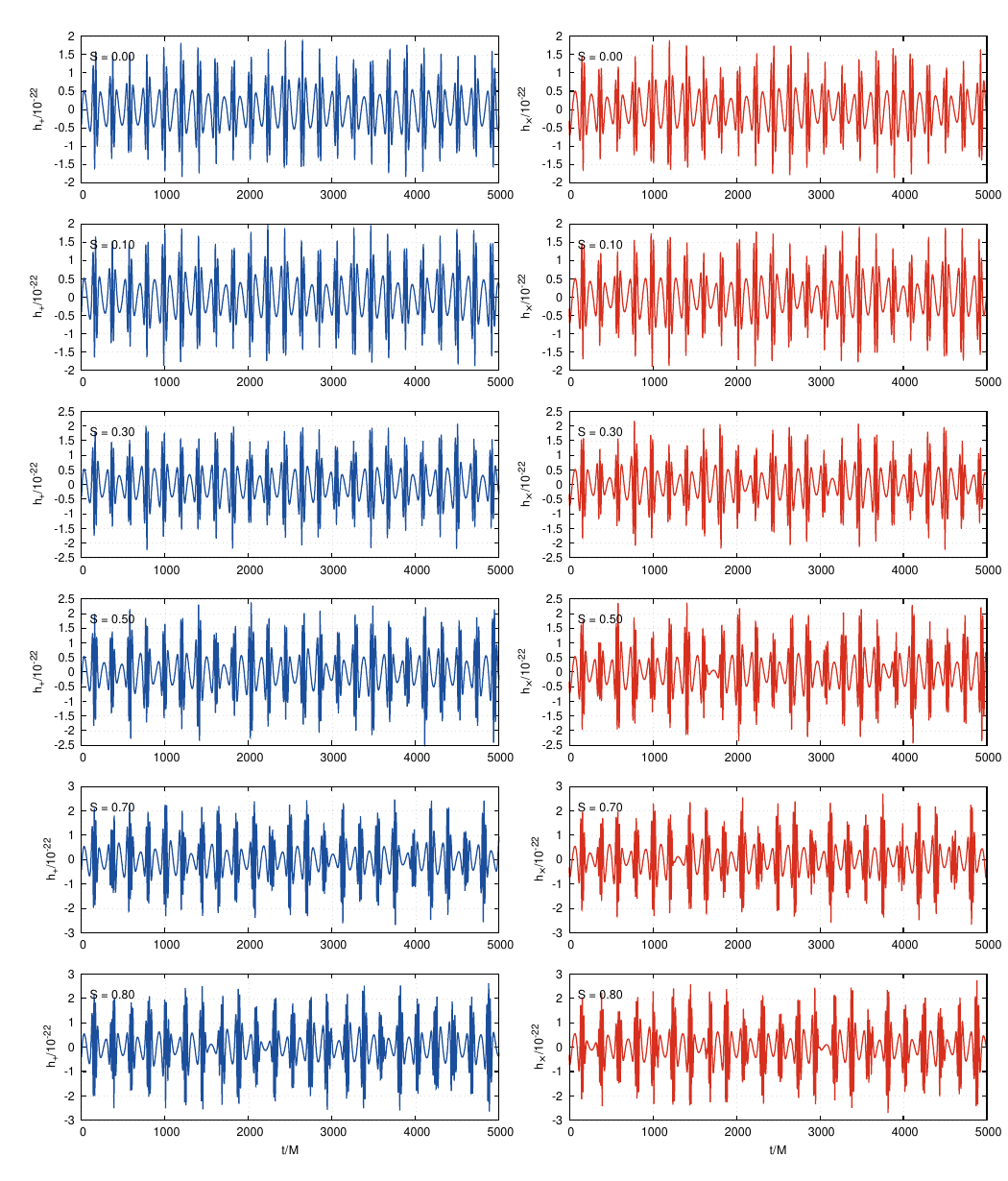}
\caption{Numerical-kludge waveforms $h_+$ (left) and $h_\times$ (right) over $5000M\simeq6.8$~h for
$(\alpha,\beta)=(0.9,0.8)$, with the spin increasing from $S=0$ (top) to $S=0.8$ (bottom). Each burst comes
from a pericenter passage, and the slow modulation of the burst heights follows the advance of the
pericenter relative to the observer. Larger spin brings the pericenter closer to the black hole and produces
stronger and narrower bursts, while the signal remains a regular burst train at every spin.}
\label{fig:time}
\end{figure*}

Figure~\ref{fig:time} displays the complete $5000M$ signal for $(\alpha,\beta)=(0.9,0.8)$. The left
column is $h_+$ and the right column is $h_\times$, and the spin increases from $S=0$ in the top row to
$S=0.8$ in the bottom row. Reading the figure row by row shows how the signal changes as the orbit
changes.

For $S=0$ the waveform is a regular train of bursts. Each burst is produced during a pericenter passage,
when the particle moves fastest and its acceleration is largest, and between bursts the signal relaxes
to a slow oscillation produced near apocenter. The bursts repeat at the radial period, and their
pattern slowly changes from one passage to the next because the pericenter advances by a large angle
on every orbit, so the orientation of the burst relative to the observer is not repeated. The cross
polarization carries the same pattern with a different projection weight.

As the spin increases, two things happen, and both can be traced to Fig.~\ref{fig:orbit}. First, the
bursts become stronger. The largest value of $|h_+|$ grows from $1.9\times10^{-22}$ at $S=0$ to
$2.5\times10^{-22}$ at $S=0.5$ and $2.6\times10^{-22}$ at $S=0.8$, because the particle passes
closer to the black hole and its acceleration at pericenter is larger. Second, the bursts become
narrower and more structured, since a deeper pericenter passage lasts a shorter time and the orbit
also precesses more strongly. The signal nevertheless stays a regular, recurring burst train at every
spin shown; there is no sign of erratic behaviour, and the envelope does not drift.

\begin{figure}[t]
\includegraphics[width=\columnwidth]{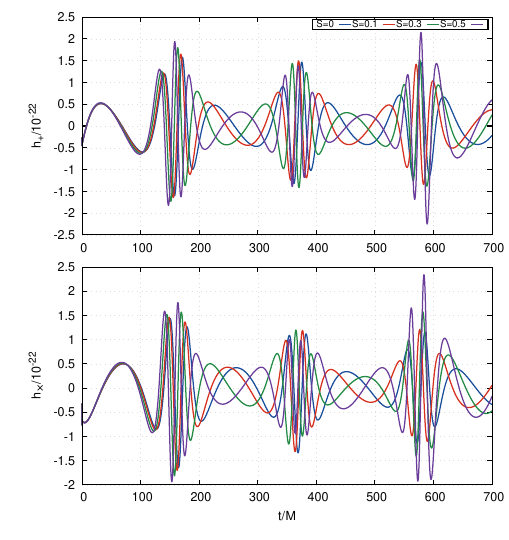}
\caption{First $700M$ of the waveforms of Fig.~\ref{fig:time} for $S=0$, $0.1$, $0.3$ and $0.5$, overlaid. Top:
$h_+$; bottom: $h_\times$. The curves coincide while the particle falls in from $r_0=8M$. The bursts keep
arriving at almost the same times, but the oscillations inside and between them go out of phase, by an
amount that grows with the spin. The leading effect of the spin is therefore an accumulated dephasing.}
\label{fig:zoom}
\end{figure}

Figure~\ref{fig:zoom} resolves the first $700M$ for $S=0$, $0.1$, $0.3$ and $0.5$. During the first
$100M$, while the particle falls inward from $r_0=8M$, the four curves lie on top of each other. The
bursts produced at the pericenter passages arrive at almost the same times for all spins, near
$t\simeq160$, $370$ and $580M$, because the spin hardly changes the radial period. What does change is
the phase of the oscillations, both inside each burst and in the slower oscillation produced near
apocenter between bursts. In the interval between the first two bursts the four curves are already
clearly out of phase, and the offset grows with the spin. This dephasing comes from the azimuthal and
precessional motion: the spin changes how far the particle advances in azimuth during each radial
cycle and slowly tilts the orbital plane, so the orientation of the orbit seen by the observer drifts
from one spin to another. At the third passage, near $t\simeq580M$, the $S=0.5$ burst is also visibly
stronger, reflecting its deeper pericenter. In a long-lived EMRI such an accumulated phase difference
matters much more than a modest change of the instantaneous amplitude.

\subsection{Frequency domain}
\label{sec:freq}

\begin{figure*}[t]
\includegraphics[width=\textwidth]{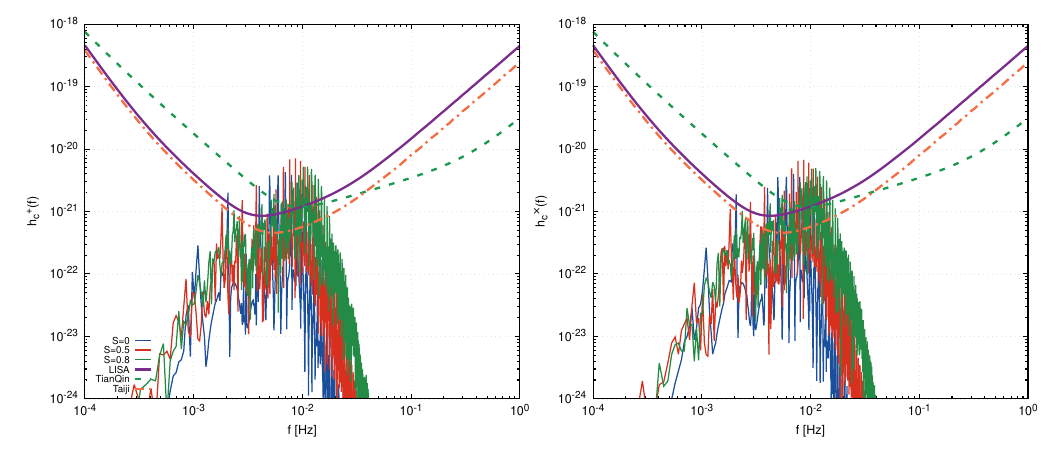}
\caption{Characteristic strain of the plus (left) and cross (right) polarizations for $(\alpha,\beta)=(0.9,0.8)$ and
$S=0$, $0.5$ and $0.8$, together with the sky-averaged sensitivity curves $h_n=\sqrt{fS_n}$ of LISA, TianQin and
Taiji. The spectrum is a comb of orbital harmonics peaking near $10^{-2}$~Hz. Increasing the spin excites higher
harmonics and moves the strongest line from about $7$ to about $11$~mHz, but no broadband component appears.
With the conventions used here $\rho^2=\int(h_c/h_n)^2\,\dd\ln f$, so the distance between the curves is a direct
measure of the signal-to-noise ratio collected at each frequency.}
\label{fig:hc}
\end{figure*}

Figure~\ref{fig:hc} shows the characteristic strain of the two polarizations for $S=0$, $0.5$ and
$0.8$, together with the sensitivity curves of LISA, TianQin and Taiji. Both panels show the same
structure. The spectrum is a comb of discrete lines, the harmonics of the orbital frequencies, which
rises from a few times $10^{-4}$~Hz to a maximum near $10^{-2}$~Hz and then falls steeply. Above about
$5\times10^{-2}$~Hz there is essentially no power. This is what one expects from a bound,
quasi-periodic orbit, and it holds for all the spins we consider.

The spin changes how the power is distributed among the harmonics. For $S=0$ the strongest harmonic
lies at about $7$~mHz. As the pericenter moves inward, higher harmonics are excited more strongly: the
comb extends to higher frequencies, and the strongest line moves to about $9$~mHz at $S=0.5$ and to
about $11$~mHz at $S=0.8$. This is the frequency-domain counterpart of the narrower and stronger bursts in
Fig.~\ref{fig:time}. Because the dominant part of the comb lies between $3$ and $20$~mHz, the signal falls
in the band where LISA and Taiji are most sensitive. TianQin, whose sensitivity is best at higher
frequencies, sees a weaker signal from this source, as we quantify in Sec.~\ref{sec:detect}.

\subsection{Linear response to the spin and realistic EMRIs}
\label{sec:linear}

The spins used so far are far larger than those of astrophysical secondaries, for which
$S\sim\chi\mu/M\lesssim10^{-5}$. To connect the two regimes we have measured how the orbital frequencies
respond to the spin. For each $S$ we evolve the reference orbit for $10^4M$ of coordinate time and compare
its mean azimuthal frequency $\Omega_\phi$ and mean radial frequency $\Omega_r$ with those of the
nonspinning orbit, together with the azimuthal phase difference $\Delta\phi$ accumulated at the end of the
window. Table~\ref{tab:linear} shows the result. The relative shift of the azimuthal frequency is linear in
$S$ to high accuracy, $\delta\Omega_\phi/\Omega_\phi\simeq0.354\,S$ in Kerr and $0.318\,S$ for
$(\alpha,\beta)=(0.9,0.8)$, with the ratio $\delta\Omega_\phi/(\Omega_\phi S)$ constant to within $2\%$ for
$S\le10^{-2}$ and $4\%$ at $S=0.03$. The radial frequency changes much less, by
$\delta\Omega_r/\Omega_r\simeq0.003\,S$ in Kerr and $0.019\,S$ in the hairy background, which is why the
bursts in Fig.~\ref{fig:zoom} keep arriving at nearly the same times while their phase drifts. The departure
from linearity is $6$--$8\%$ at $S=0.1$ and $50$--$70\%$ at $S=0.5$, so the larger spins probe the onset of
the nonlinear regime while still following the same mechanism.

For an astrophysical secondary the same mechanism shifts the azimuthal frequency by
$\delta\Omega_\phi/\Omega_\phi\sim3\times10^{-6}$ at $S=10^{-5}$. Over a $10^4M\simeq14$~h window this gives only
$\Delta\phi\simeq2\times10^{-3}$~rad, but over the $\sim10^5$ cycles of an inspiral it accumulates to a phase of
order $2\pi\times10^5\times3\times10^{-6}\sim1$~rad, consistent with the known result that the spin of the
secondary enters the phase of an EMRI at first post-adiabatic order, producing phase corrections of order
unity over a complete inspiral~\cite{HindererFlanagan2008,Barack2019}. The large-$S$ results of this paper
should therefore be read as a magnified illustration of this physical mechanism, not as a direct
astrophysical prediction.

\begin{table}[t]
\caption{Spin-induced shift of the azimuthal frequency, $\delta\Omega_\phi/(\Omega_\phi S)$, and of the azimuthal
phase accumulated over $10^4M$, $\Delta\phi/S$ (rad), for the reference orbit $r_p=4M$, $e=0.5$. Both ratios
are constant at small $S$, showing that the response is linear in the spin.}
\label{tab:linear}
\begin{ruledtabular}
\begin{tabular}{lcccc}
 & \multicolumn{2}{c}{Kerr} & \multicolumn{2}{c}{$(\alpha,\beta)=(0.9,0.8)$} \\
$S$ & $\delta\Omega_\phi/(\Omega_\phi S)$ & $\Delta\phi/S$ & $\delta\Omega_\phi/(\Omega_\phi S)$ & $\Delta\phi/S$ \\
\colrule
$10^{-4}$ & 0.3535 & 167.8 & 0.3175 & 149.2 \\
$10^{-3}$ & 0.3537 & 167.9 & 0.3171 & 149.0 \\
$10^{-2}$ & 0.3602 & 171.0 & 0.3140 & 147.5 \\
$0.03$    & 0.3666 & 174.0 & 0.3206 & 150.7 \\
$0.1$     & 0.3832 & 181.9 & 0.3360 & 157.9 \\
$0.3$     & 0.4357 & 206.8 & 0.3918 & 184.1 \\
$0.5$     & 0.5879 & 279.0 & 0.4818 & 226.4 \\
\end{tabular}
\end{ruledtabular}
\end{table}

\subsection{What part of the signal comes from the hair?}
\label{sec:hairwave}

\begin{figure*}[t]
\includegraphics[width=\textwidth]{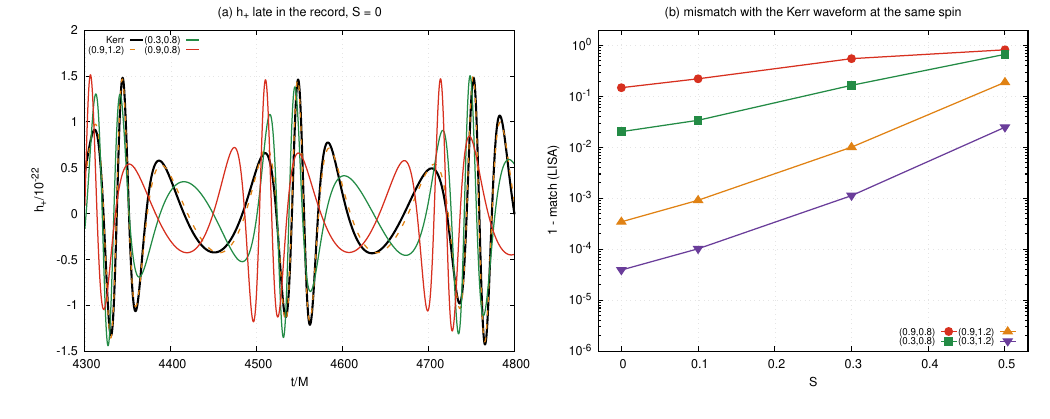}
\caption{Imprint of the hair on the waveform. (a) $h_+$ for a nonspinning particle in four backgrounds, in a
window near the end of the $5000M$ record. The $(0.9,1.2)$ waveform coincides with Kerr, the $(0.3,0.8)$ waveform
is slightly shifted and the $(0.9,0.8)$ waveform is clearly out of phase; the burst shapes are nearly the same.
(b) Mismatch between the hairy waveform and the Kerr waveform with the same spin, weighted by the LISA noise
and maximized over time and phase. The mismatch is controlled mainly by $\beta$, and it grows with the spin because
a spinning particle samples more of the deformed region. For $\beta=1.6$ the mismatch stays below $10^{-6}$ and lies outside the plot. Mismatches well below the
few-percent accuracy of the NK model only indicate that the effect is small (see text).}
\label{fig:hair}
\end{figure*}

Figure~\ref{fig:hair}(a) isolates the effect of the hair. It shows $h_+$ for a nonspinning particle in
four backgrounds, in a window near the end of the record where differences have had time to build up.
The waveform for $(\alpha,\beta)=(0.9,1.2)$ lies exactly on top of the Kerr waveform, the waveform for
$(0.3,0.8)$ is shifted slightly, and the waveform for $(0.9,0.8)$ is clearly out of phase with Kerr. The
shapes of the individual bursts are very similar in all four cases; what the hair changes is when they
arrive. This is the same mechanism as for the spin: the hair slightly changes the orbital frequencies,
and the difference accumulates into a phase shift.

To quantify the effect we compute the overlap between the hairy waveform and the Kerr waveform with
the same spin, weighted by the LISA noise and maximized over the time and phase of arrival, as defined
in Sec.~\ref{sec:detect}. Figure~\ref{fig:hair}(b) shows the mismatch, one minus this overlap. Three
features stand out. First, $\beta$ matters far more than $\alpha$. For a nonspinning particle the
mismatch is $0.15$ for $(0.9,0.8)$ but only $3.5\times10^{-4}$ for $(0.9,1.2)$, a reduction by a factor of
about 400 for a modest change of $\beta$; reducing $\alpha$ from $0.9$ to $0.3$ at fixed $\beta$ lowers it by
a factor of about 7--9. For $\beta=1.6$ the waveforms are identical to the Kerr waveforms to numerical
precision, with mismatches below $10^{-6}$ at all spins, so they do not appear in the figure. This is the
behaviour anticipated in Sec.~\ref{sec:metric}: $\alpha$ only sets the size of the deformation, while
$\beta$ decides whether the orbit reaches the region where the deformation is present at all. Second, the
mismatch grows rapidly with the spin. For $(0.9,0.8)$ it rises from $0.15$ at $S=0$ to $0.82$ at $S=0.5$,
and for $(0.9,1.2)$ from $3.5\times10^{-4}$ to $0.19$. A spinning particle dips deeper into the strong-field
region, where the deformation is largest, so the spin amplifies the imprint of the hair. Third,
the ordering of the four curves never changes, which shows that the hair signal is controlled by
the same combination of $\alpha$ and $\beta$ at every spin.

The nonspinning results deserve emphasis, because they do not rely on the large spins used elsewhere in
this paper. For an astrophysical secondary, $S\lesssim10^{-5}$, the spin shifts the orbital phase by less
than $10^{-3}$~rad over the $5000M$ window (Sec.~\ref{sec:linear}), so it is effectively zero here, and the
$S=0$ results of Fig.~\ref{fig:hair} describe the hair imprint of a realistic EMRI directly, within the
conservative numerical-kludge approximation used throughout. The growth of the imprint with $S$, on the
other hand, is a magnified illustration of how a spinning secondary samples more of the deformed region.

\subsection{Detectability and distinguishability}
\label{sec:detect}

The sensitivity curves in Fig.~\ref{fig:hc} indicate the frequency band of the signal, but whether
the signal can be detected has to be decided by its signal-to-noise ratio. For a detector with
one-sided noise spectral density $S_n(f)$ we use the noise-weighted inner product~\cite{Finn1992,CutlerFlanagan1994}
\begin{equation}
(a|b) = 4\,{\rm Re}\int_{f_{\min}}^{f_{\max}}
\frac{\tilde a_+(f)\tilde b_+^*(f)+\tilde a_\times(f)\tilde b_\times^*(f)}{S_n(f)}\,\dd f ,
\label{eq:inner}
\end{equation}
with $f_{\min}=10^{-4}$~Hz and $f_{\max}=1$~Hz, and the signal-to-noise ratio $\rho=\sqrt{(h|h)}$. For
LISA we take the sky-averaged sensitivity of Robson, Cornish and Liu~\cite{RCL2019}, for Taiji the same
form with an arm length of $3\times10^6$~km and an optical-metrology noise of
$8$~pm\,Hz$^{-1/2}$~\cite{HuWu2017}, and for TianQin the sky-averaged form with an arm length of
$\sqrt3\times10^5$~km~\cite{Luo2016,Li2025}, as in Ref.~\cite{Das2025b}. With these conventions
$\rho^2=\int(h_c/h_n)^2\,\dd\ln f$ with $h_n=\sqrt{fS_n}$, so the visual comparison in Fig.~\ref{fig:hc} and
the numbers below are consistent.

\begin{figure*}[t]
\includegraphics[width=\textwidth]{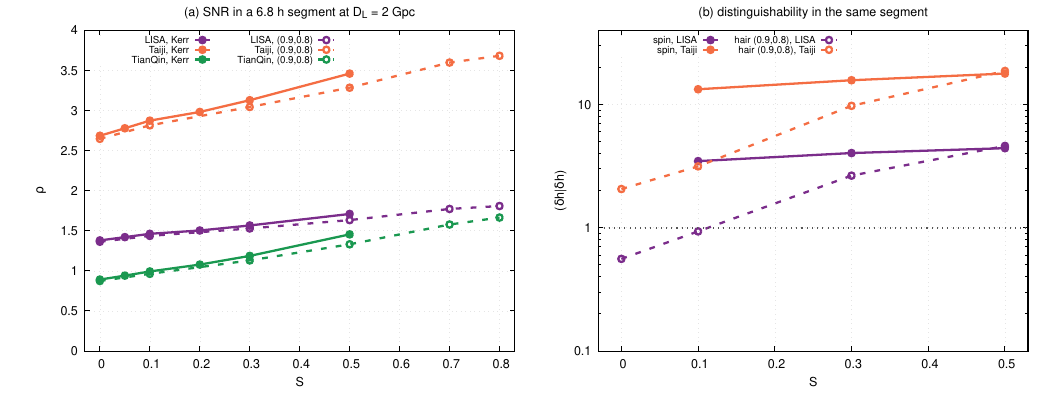}
\caption{Observability of the signal from a source at $2$~Gpc in a $6.8$~h segment. (a) Signal-to-noise ratio
in LISA, Taiji and TianQin as a function of the spin, for Kerr (solid) and for $(\alpha,\beta)=(0.9,0.8)$ (dashed).
Taiji collects the largest signal because the harmonics lie where its noise is lowest; the signal grows with
the spin as the particle comes closer to the black hole. (b) Distinguishability $(\delta h|\delta h)$ of the
spinning waveform from the nonspinning one in Kerr (solid), and of the $(0.9,0.8)$ waveform from the Kerr waveform
at the same spin (dashed), for LISA and Taiji. Values above the dotted line, $(\delta h|\delta h)=1$, can be told apart.}
\label{fig:obs}
\end{figure*}

Figure~\ref{fig:obs}(a) and Table~\ref{tab:snr} give the signal-to-noise ratio accumulated in the
$6.8$~h window for a source at $2$~Gpc. Taiji performs best, with $\rho\simeq2.7$ for a nonspinning particle,
followed by LISA with $\rho\simeq1.4$ and TianQin with $\rho\simeq0.9$. The ordering follows from
Fig.~\ref{fig:hc}: the signal peaks between $5$ and $10$~mHz, where Taiji has the lowest noise. The
signal-to-noise ratio grows with the spin, by about $25\%$ at $S=0.5$ and by about $35\%$ at $S=0.8$, because
the particle comes closer to the black hole and radiates more strongly. The hair changes $\rho$ by only a
few percent, so detection by itself carries little information about the hair; that information is in
the phase.

\begin{table}[t]
\caption{Signal-to-noise ratio $\rho$ accumulated in a $5000M\simeq6.8$~h window for a source at $2$~Gpc,
and the time $T_{20}$ needed to reach $\rho=20$ if the orbit did not evolve, $T_{20}=6.8~{\rm h}\times(20/\rho)^2$.}
\label{tab:snr}
\begin{ruledtabular}
\begin{tabular}{lccccccc}
 & & \multicolumn{3}{c}{$\rho$ in $6.8$~h} & \multicolumn{3}{c}{$T_{20}$ [days]} \\
Background & $S$ & LISA & Taiji & TianQin & LISA & Taiji & TianQin \\
\colrule
Kerr & 0   & 1.38 & 2.68 & 0.89 & 60 & 16 & 144 \\
Kerr & 0.5 & 1.71 & 3.46 & 1.45 & 39 & 10 & 54 \\
$(0.9,0.8)$ & 0   & 1.36 & 2.65 & 0.87 & 61 & 16 & 149 \\
$(0.9,0.8)$ & 0.5 & 1.63 & 3.28 & 1.33 & 43 & 11 & 64 \\
$(0.9,0.8)$ & 0.8 & 1.81 & 3.68 & 1.66 & 35 & 8 & 41 \\
\end{tabular}
\end{ruledtabular}
\end{table}

A signal-to-noise ratio of a few in under seven hours is of course not a detection. Real EMRI searches
integrate for months, and a threshold of $\rho=20$ is commonly adopted~\cite{Babak2017}. For a signal of
fixed frequency content $\rho$ grows as the square root of the observation time, so Table~\ref{tab:snr}
also lists the time $T_{20}$ needed to reach this threshold. Taiji reaches it within about two weeks and
LISA within one to two months, both shorter than the radiation-reaction time of about three months. For
TianQin the required time exceeds $T_{\rm rr}$, so the orbit would evolve appreciably before detection
and a full inspiral model would be needed to decide the question. These estimates should be read as
indicative, since they rest on the kludge amplitude and on a fixed observer orientation.

Detecting a signal and telling two signals apart are different questions. Two waveforms $h_1$ and $h_2$
can be distinguished at roughly the one-sigma level when $(\delta h|\delta h)\ge1$, with
$\delta h=h_1-h_2$~\cite{Lindblom2008}. We evaluate this after maximizing the overlap over a relative time
and phase shift, so that only genuine differences in the signal are counted. Figure~\ref{fig:obs}(b)
shows the result. For the spins used here the secondary spin is easily
distinguishable: already for $S=0.1$ one has $(\delta h|\delta h)\simeq3.5$ in LISA and $13$ in Taiji within the
$6.8$~h window, and the overlap between the $S=0.1$ and nonspinning waveforms is only $0.14$. For an
astrophysical spin, $S\sim10^{-5}$, the linear scaling of Sec.~\ref{sec:linear} reduces the dephasing to
below $10^{-3}$~rad over this window, so the spin of a real secondary cannot be distinguished within hours;
its effect becomes important only as it accumulates over a long inspiral. The hair is harder to see. For a nonspinning
particle in the $(0.9,0.8)$ background, $(\delta h|\delta h)=0.56$ in LISA, just below the threshold, and
$2.1$ in Taiji, just above it. Because the two waveforms dephase steadily, $(\delta h|\delta h)$ grows with the
observation time, and LISA would cross the threshold within about half a day. For $\beta=1.2$ and
nonspinning motion the same quantity is only $1.3\times10^{-3}$ in LISA, and for $\beta=1.6$ the hair is
invisible. Since an astrophysical secondary spin is effectively zero on these timescales, these
nonspinning values are the ones that apply directly to a realistic EMRI. As in Fig.~\ref{fig:hair}(b), spin makes the hair easier to see: at $S=0.5$ the $(0.9,0.8)$ background
gives $(\delta h|\delta h)=4.6$ in LISA and $19$ in Taiji.

Since both the spin and the hair act through the phase, one may ask whether one can mimic the
other. We have tested the simplest version of this question by comparing the nonspinning waveform in
the $(0.9,0.8)$ background with Kerr waveforms of increasing spin. The best overlap, $0.85$, is obtained
for a nonspinning Kerr particle, and adding spin to the Kerr particle makes the agreement worse, with
overlaps of $0.26$, $0.15$ and $0.05$ for $S=0.05$, $0.1$ and $0.3$. Within this comparison the hair does not
look like a secondary spin. The reason is that the two act on the orbit differently: the spin mainly
deepens the pericenter, while the hair changes the frequencies without moving the turning points. A complete answer would require a parameter-estimation study that also varies the masses and the orbital elements, which is beyond the scope of this work.

These numbers have to be read with the accuracy of the waveform model in mind. In Kerr, numerical-kludge
waveforms reproduce Teukolsky-based ones with overlaps of about $95\%$~\cite{Babak2007}, so the model itself
carries a systematic mismatch of a few percent, mostly in the amplitudes and in the relative weights of the
harmonics. Mismatches between two NK waveforms that lie well above this level, such as those produced by
the secondary spin ($1-\mathcal{M}\gtrsim0.86$ for $S\ge0.1$) or by the hair with $\beta=0.8$ ($0.15$ at $S=0$
and $0.82$ at $S=0.5$), come from accumulated dephasing. That dephasing is controlled by the orbital
frequencies, which the NK construction inherits exactly from the relativistic trajectory, so these
conclusions are robust. Mismatches far below the model accuracy, such as $3.5\times10^{-4}$ for $(0.9,1.2)$ or
the values below $10^{-6}$ for $\beta=1.6$, should not be overinterpreted. They show that the effect is small
and that the Kerr limit is recovered, but their precise values are not predictions for real signals, and the
corresponding values $(\delta h|\delta h)\ll1$ only mean that such deformations cannot be resolved with this
model over the window considered.

\section{Physical interpretation and limitations}
\label{sec:disc}

The results can be summarized in terms of three pieces of physics. First, the spin changes the orbit.
Through the MPD force it pulls the pericenter inward, strengthens and sharpens the bursts emitted there,
shifts the orbital frequencies and, beyond a threshold, carries the particle into the black hole.
Second, the hair changes the geometry on which the spinning particle moves. Because the deformation is
localized in radius, its effect depends on where the orbit goes: it is decided by $\beta$, which sets the
reach of the deformation, much more than by $\alpha$, which sets its size. Near the black hole the hair
reduces the effective mass, which raises the spin needed for capture. Third, the waveform translates
these orbital changes into a phase shift between waveforms and into a redistribution of power among the
harmonics of the orbit. Throughout the parameter range we consider, the signal remains a discrete comb of
harmonics; neither the spin nor the hair produces a broadband component.

The two effects also act together. The deeper a spinning particle dips, the more of the deformed
region it samples, so the waveform imprint of the hair grows with the spin, and the capture threshold
is most strongly shifted where the deformation is largest. This is the waveform-level counterpart of the
cooperative spin--hair behaviour found for the finite-time instability in our companion
paper~\cite{DaluiGe2026}. The regular line spectra found here at $r_p=4M$ agree with Hartl's finding that
generic orbits in Kerr show no chaos for small spins~\cite{Hartl2003a}, while Yuan \emph{et al.}
have shown that orbits close to the separatrix can be chaotic even for astrophysically realistic
secondary spins~\cite{Yuan2026}. Any chaotic
imprint on the waveform would have to be sought close to the capture threshold, where orbits linger near
the separatrix before plunging; resonant effects in non-integrable spacetimes are known to be most
pronounced in such regions~\cite{Apostolatos2009}. Combining the waveform and the Lyapunov analyses for the same
trajectories near the threshold is a natural next step.

Several limitations should be kept in mind. The NK construction is approximate and cannot replace perturbation theory in the hairy background, especially
for absolute amplitudes; mismatches below its few-percent accuracy should not be overinterpreted
(Sec.~\ref{sec:detect}). Radiation reaction is
neglected, so the signals describe conservative orbits over a window much shorter than the inspiral.
The spins we use are much larger than those of astrophysical secondaries, for which
$S\sim\chi\mu/M\lesssim10^{-5}$. As shown in Sec.~\ref{sec:linear}, the spin-induced frequency shifts are linear
in $S$ at small spin, so a realistic secondary spin produces a negligible dephasing within a few hours but a
phase of order a radian over a complete inspiral, which is why it has to be included in waveform models for
tests of the Kerr geometry~\cite{HindererFlanagan2008,Barack2019}. The large values of $S$ used here are a
magnifying glass on this effect. Finally, the distinguishability estimates use a fixed observer orientation and do not
explore correlations with the other source parameters.

\section{Conclusions}
\label{sec:concl}

We have studied numerical-kludge gravitational waves from a spinning compact secondary on an eccentric,
inclined orbit around a rotating hairy black hole obtained by gravitational decoupling. The
deformation admits a black hole only below a critical curve in the plane of the hair parameters, and
this geometric boundary is the natural place to look for the strongest imprint of the hair.

Our main conclusions are the following. The spin of the secondary moves the pericenter inward and,
above a critical value, carries the particle into the black hole. The hair raises this capture
threshold, from $S_{\rm crit}\simeq0.66$ in Kerr to $0.9$ and above next to the horizon boundary, because the
deformation reduces the effective mass close to the black hole. The emitted waves remain a discrete
superposition of orbital harmonics for all the spins we consider. The spin strengthens the bursts and
shifts power to higher harmonics, and both the spin and the hair act mainly through an accumulated
dephasing. The imprint of the hair is controlled by the parameter $\beta$ that sets the radial reach of
the deformation, it vanishes in the Kerr-like limit of the family, and it is amplified by the spin of
the secondary.

For a source at $2$~Gpc with $M=10^6M_\odot$ and $\mu=10M_\odot$, the signal lies in the band where LISA and
Taiji are most sensitive. In a $6.8$~h segment it carries $\rho\simeq1.4$ in LISA, $2.7$ in Taiji and $0.9$ in
TianQin, so Taiji and LISA would reach a detection threshold of $\rho=20$ within about two weeks and two
months, respectively. Within the same segment the hair becomes
distinguishable for $\beta\lesssim0.8$, first in Taiji and then in LISA; since this result holds for a
nonspinning secondary, it applies directly to astrophysical EMRIs, whereas the clear distinguishability of
the large spins used here is a magnified illustration.

The spin-induced frequency shifts are linear in $S$ in the small-spin regime, so the large spins used here
are a magnified illustration of a mechanism that, for astrophysical spins, accumulates to a phase of order a
radian over an inspiral; the large dephasing effects are robust against the few-percent accuracy of the NK
model, whereas very small mismatches should not be overinterpreted.

These results suggest that tests of non-Kerr structure with extreme-mass-ratio systems should treat
the spin of the secondary and the geometry of the central object together, since the spin both shifts
the phase of the signal and controls how deeply the orbit probes the deformation. The most important
extensions are the inclusion of radiation reaction, which would turn these conservative signals into
inspirals, and a parameter-estimation study of the correlations between the hair, the secondary spin and
the other source parameters.

\begin{acknowledgments}
This work was partially supported by the National Natural Science Foundation of China (NSFC) under Grant
Nos.~12275166 and 12675074. 
\end{acknowledgments}

\appendix

\section{Numerical evolution and checks}
\label{app:num}

We integrate the MPD system in terms of the momentum $p_\mu$ and the spin tensor $S^{\mu\nu}$, using
Eq.~\eqref{eq:mpd1} for the force, Eq.~\eqref{eq:mpd2} for the spin tensor and the exact relation
\eqref{eq:vp} for the velocity, with a fourth-order Runge--Kutta scheme in proper time and a step
$\Delta\tau=0.0125M$. When the equations are written in terms of the spin vector instead, the force and
the evolution of the spin vector involve the right dual of the Riemann tensor, and the velocity relation
involves its double dual~\cite{Hartl2003a}. Working with $S^{\mu\nu}$ avoids dual tensors altogether. It also
makes the conservation of $E$ and $J_z$ a sharp test of the whole implementation, because these
quantities are conserved by Eqs.~\eqref{eq:mpd1} and~\eqref{eq:mpd2} for any choice of velocity relation
but only if the force is computed correctly. We have checked that replacing the dual by the Riemann
tensor itself in the spin-vector form leads to large violations of these conservation laws, of order
unity in $J_z$ for $S=1$ even in Kerr, which is a useful diagnostic for codes that use the spin-vector
form.

Table~\ref{tab:checks} summarizes the checks. In all bound runs at $r_p=4M$ the energy and angular
momentum are conserved to better than $10^{-10}$ and $2\times10^{-9}$, the mass shell and the Tulczyjew
condition are satisfied to better than $3\times10^{-10}$, and the spin magnitude is conserved to the same
level. No projection onto the constraint surface is needed. The signal-to-noise ratios and the turning
points are unchanged to the quoted digits when the step is changed from $0.025M$ to $0.00625M$. For $S=0$
the evolution reproduces the geodesics of the background, and in the Kerr limit the horizon is
recovered to machine precision.

\begin{table}[t]
\caption{Checks of the numerical evolution for $r_p=4M$ over $5000M$ of proper time. $\Delta E$ and $\Delta J_z$
are the largest deviations from the initial values. Changing the step from $0.025M$ to $0.00625M$ leaves
the LISA signal-to-noise ratio unchanged to five digits ($1.7091$ for Kerr and $1.6325$ for $(0.9,0.8)$ at $S=0.5$).}
\label{tab:checks}
\begin{ruledtabular}
\begin{tabular}{lcccc}
Background & $S$ & $\Delta E$ & $\Delta J_z$ & $|p^\mu p_\mu+1|$ \\
\colrule
Kerr        & 0.5 & $2\times10^{-11}$ & $6\times10^{-10}$ & $5\times10^{-11}$ \\
$(0.9,0.8)$ & 0.5 & $2\times10^{-11}$ & $7\times10^{-10}$ & $5\times10^{-11}$ \\
$(0.9,0.8)$ & 0.8 & $4\times10^{-11}$ & $1\times10^{-9}$  & $3\times10^{-10}$ \\
\end{tabular}
\end{ruledtabular}
\end{table}

The waveform is computed by resampling the trajectory on a uniform grid in Boyer--Lindquist time with
spacing $0.05M$ and differentiating with cubic splines. The spectra are evaluated on the exact
frequency grid of the record. The overlaps of Sec.~\ref{sec:detect} are computed from zero-padded records,
so that the maximization over the time shift is not affected by the periodicity of the discrete Fourier
transform.

\bibliography{Hairy_GW_PRD}

\end{document}